\documentclass[prd,eqsecnum,preprint,showpacs,nofootinbib]{revtex4}
\usepackage{amsthm}
\usepackage{amssymb}
\usepackage{amsmath}

\usepackage{graphicx}
\usepackage[caption=false]{subfig}

\newcommand{\be}{\begin{equation}}
\newcommand{\ee}{\end{equation}}
\newcommand{\bea}{\begin{eqnarray}}
\newcommand{\eea}{\end{eqnarray}}
\newcommand{\nn}{\nonumber}

\begin{document}

\title{Scalar Casimir Energies of Tetrahedra and Prisms}
\author{E. K. Abalo}
\email{abalo@nhn.ou.edu}
\author{K. A. Milton}
\email{milton@nhn.ou.edu}
\affiliation{Homer L. Dodge Department of Physics and Astronomy, 
University of Oklahoma, Norman, OK 73019}
\author{L. Kaplan}
\email{lkaplan@tulane.edu}
\affiliation{Department of Physics, Tulane University, New Orleans,
LA 70118}
\date{\today}


\begin{abstract}
New results for scalar Casimir self-energies arising from interior modes are presented for the 
three integrable tetrahedral cavities. Since the eigenmodes are all known, the energies
can be directly evaluated by mode summation, with a point-splitting regulator, which amounts
to evaluation of the cylinder kernel.  The correct Weyl divergences, depending on the volume,
surface area, and the edges, are obtained, which is strong evidence that the counting
of modes is correct.  Because there is no curvature, the finite part of the quantum energy
may be unambiguously extracted. Cubic, rectangular parallelepipedal, triangular prismatic, and 
spherical geometries are also revisited. Dirichlet and Neumann boundary conditions are considered 
for all geometries.
Systematic behavior of the energy in terms of geometric invariants for 
these different cavities is explored.  Smooth interpolation between short
and long prisms is further demonstrated.  When scaled by the ratio of the 
volume to the surface area, the energies for the tetrahedra and the prisms
of maximal isoareal quotient lie very close to a universal curve. 
The physical significance of these results is discussed.
\end{abstract}

\maketitle

\section{Introduction}
The concept of Casimir self-energy remains elusive. 
Since 1948, the year of H. B. G. Casimir's seminal paper \cite{casimir}, what is now 
called the Casimir effect has captivated many. Casimir discovered an attractive 
quantum vacuum force between uncharged parallel conducting plates.
Yet, while Casimir later extrapolated from this to predict an 
attractive force for a spherical conducting shell \cite{casimir53},
Boyer proved the self-stress in that case to be instead repulsive,
 which was an even more unexpected result \cite{boyer}. 
Since Boyer's formidable calculation, many other configurations were examined: 
cylinders, boxes, wedges, etc. The literature abounds with these results;
for a review see Ref.~\cite{milton-rev}.  However, since there are other
well-known cases of cavities where the interior modes are known exactly,
it is surprising that essentially no attention had been paid to these.
For example, recently we presented the first results for Casimir self-energies
for cylinders of equilateral, hemiequilateral, and right isosceles triangular
cross sections \cite{Cylpaper}, even though the spectrum is well-known and appears in general
textbooks \cite{radbook,embook}.  
Possibly, the reason for this neglect was that only interior modes could 
be included for any of these cases, unlike the case of
a circular cylinder, where both interior and exterior modes must be included in
order to obtain a finite self-energy.  However, the extensive attention to
rectangular cavities puts the lie to this hypothesis \cite{lukosz,
lukosz2,lukosz3,ambjorn,ruggiero1,ruggiero2}.
It seems not to have been generally appreciated that finite results can be obtained
in all these cases because there are no curvature divergences for boxes constructed
from plane surfaces.

In this paper, as in Ref.~\cite{Cylpaper}, we put aside the serious objection that
these self-energies may be impossible to observe, even in principle.
\footnote{The exception would be in the coupling to gravity.  Since it
is highly likely that Casimir energies obey the equivalence 
principle \cite{Fulling:2007xa},
we expect that like any other contribution to the self-energy of a body,
the Casimir energy would contribute to the inertial and gravitational
mass of a body \cite{Shajesh:2007sc,Milton:2007ar} .}
  For example,
the positive self-energy of a spherical shell is not the negative of the work 
required to separate two hemispheres, which must be positive.
We also are unable to comment on the exterior contributions to the
Casimir energy, which would be extremely difficult to calculate for any
of these boxes, since the Helmholtz equation is not separable exterior
to any box with flat sides. Nonetheless, except for geometries with smooth boundaries, one would expect an interesting progression solely for interior energies.
The fact that a smooth uniform behavior is observed
suggests that a physical/mathematical significance lies here.  Also
the interior Casimir energy can be relevant to physical situations; for
example, the interior zero-point energy of gluons is of crucial significance
for the bag model of hadrons, where the fields exist only inside the cavity
\cite{Milton:1980ke}.

In Ref.~\cite{Cylpaper} we obtained exact, closed-form results for the three above-mentioned
integrable triangles, both in a plane, and for cylinders with the corresponding cross section,
for Dirichlet, Neumann, and perfect conducting boundary conditions.
The expected Weyl divergences related to the area, perimeter, and the corners of the
triangles were obtained, going a long way toward verifying the counting of modes, which
is the most difficult aspect of these calculations.  Moreover, we were able to successfully
interpolate between the results for these triangles by using an efficient numerical evaluation,
and showed that the energies 
lie on a smooth curve, which was reasonably well-approximated by the result of a proximity
force calculation.  In this paper we show that the same techniques can be applied to 
tetrahedral boxes; again, there are exactly three integrable cases, where an explicit
spectrum can be written down. Again, it is surprising that the Casimir energies for
these cases are not well-known.  The only treatment of a pyramidal box found in the Casimir energy 
literature appears in a relatively unknown work of Ahmedov and Duru \cite{AhmedovSmT}, which,
however, seems to contain a counting error.

In this paper we present Casimir energy calculations for the three integrable tetrahedra. For each cavity we consider a massless scalar field subject to Dirichlet and Neumann boundary conditions
on the surfaces. We regulate the mode summation by temporal point-splitting, which amounts to 
evaluation of what is called the cylinder kernel \cite{Fulling:2003zx}, and extract both divergent
(as the regulator goes to zero) and finite contributions to the energy.
We also revisit cubic, rectangular parallelepipedal, triangular prismatic, and spherical 
geometries with the same boundary conditions. In the end, we explore the functional behavior 
of the Casimir energies with respect to an appropriately chosen ratio of the cavities' 
volumes and surface areas.  We also examine limits as  the prism length tends to zero and
infinity, which correspond to the Casimir parallel plate and infinite cylinder limits, respectively.
	
\subsection{Point-splitting regularization}
We regularize our results by temporal point-splitting. As explained in 
Ref.~\cite{Cylpaper}, after a Euclidean rotation, we obtain
\begin{equation}
E=\frac{1}{2}\lim_{\tau \rightarrow 0}\left(-\frac{d}{d\tau}\right)\sum_{kmn}
 e^{-\tau \sqrt{\lambda_{kmn}^2}}\,\,,
\end{equation}
where the sum is over the quantum numbers that characterize the eigenvalues, and $\tau$ is
the Euclidean time-splitting parameter, supposed to tend to zero at the end of the calculation.
One recognizes the sum as the traced cylinder kernel \cite{Fulling:2003zx}.
Next, we proceed to re-express the sum with Poisson's summation formula.
\subsection{Poisson resummation}
Poisson's summation formula allows one to recast a slowly convergent sum into a more rapidly 
convergent sum of its Fourier transform,
\begin{equation}
 \sum_{m=-\infty}^\infty f(m)=\sum_{n=-\infty}^\infty \left(\int_{-\infty}^{\infty}e^{2 \pi i m n}f(m)\,dm\,\right).
\label{psf}
\end{equation}
By point-splitting and resumming, we are able to isolate the finite parts, which are Casimir self-energies,
and the corresponding divergent parts, which are the Weyl terms (see Appendix A for more detail).

\section{Casimir energies of Tetrahedra}
The three integrable tetrahedra mentioned above are not recent discoveries. 
They have, in fact, been the subject of a few articles \cite{Krishnamurthy, Turner, TerrasSwanson}. 
However, there appears to be only one Casimir energy article concerning one of these tetrahedra, 
which we denote as the ``small" tetrahedron \cite{AhmedovSmT}. These tetrahedra are integrable
in the sense that their eigenvalue spectra are known explicitly, and there are no other such
tetrahedra. We will successively look at the ``large," ``medium," and ``small" tetrahedra, as
defined below, and obtain interior scalar Casimir energies for Dirichlet and Neumann boundary conditions.
Although the exterior problems cannot be solved in these cases, 
the finite parts of the  interior energies are well defined
because the curvature is zero, and hence the second heat kernel coefficient vanishes.
\subsection{Large Tetrahedron}
\begin{figure}[h]	
	\centering
\includegraphics[scale=1, trim= 5cm 10cm 5cm 10cm, clip]{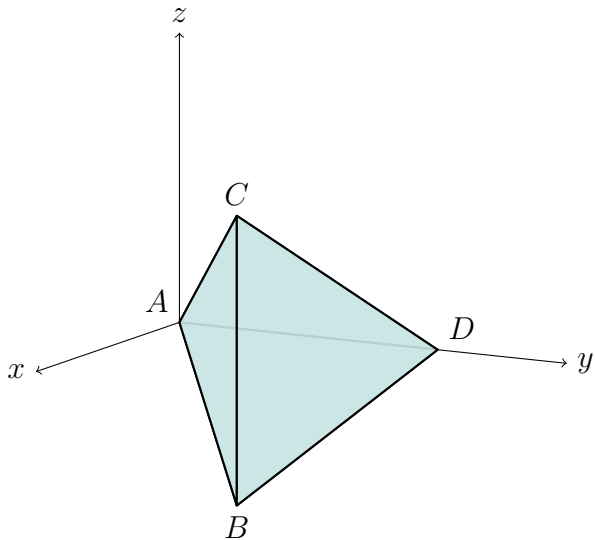}
	\caption{Large tetrahedron: $-x<z<x$ and $x<y<2a-x$. }
	\label{LargeTfigure}
	\end{figure}
The first tetrahedron, sketched in Fig.~\ref{LargeTfigure},
 which we denote ``large," is comparatively the largest or rather the most symmetrical. 
One can obtain a medium tetrahedron by bisecting a large tetrahedron and idem for the 
small and medium tetrahedra. One should note that the terms ``large,"
 ``medium," and ``small''
are merely labels, since one can always rescale each tetrahedron independently 
of the others. The spectrum and complete eigenfunction set for the large 
tetrahedron, as well as those of the other tetrahedra, are known and appear in Ref.~\cite{TerrasSwanson},
\be \label{LTeigval}
\lambda_{kmn}^2=\frac{\pi^2}{4a^2}\left[3(k^2+m^2+n^2)-2(km+kn+mn)\right]\,.
\ee
With Dirichlet and Neumann boundary conditions, different constraints are imposed on the spectrum,
that is, on the ranges of the integers $k$, $m$, and $n$.
		
\subsubsection{Dirichlet BC}
The complete set of modes for Dirichlet boundary conditions is given by the restrictions $0<k<m<n$. 
After extending the sums to all of $(k,m,n)$-space, we remove all the unphysical cases which are $k=0$, $m=0$, $n=0$, $k=m$, $k=n$, and $m=n$ while keeping track of the 24 degeneracies and compensating for oversubtractions. Finally, after suitable redefinitions of some individual terms, the 
Dirichlet Casimir energy for the large tetrahedron can be defined in terms of the function
\be \label{gfunction}
g(p,q,r)=e^{-\tau
\sqrt{(\pi/a)^2\left(p^2+q^2+r^2\right)}}\,,
\ee
and written as
\begin{align}
E=\frac{1}{48}\lim_{\tau \rightarrow 0}\left(-\frac{d}{d\tau}\right)\sum_{p,q,r=-\infty}^{\infty}
\Big[ & g(p,q,r)+g(p+1/2,q+1/2,r+1/2)-6\,g(p,q,q)
\\& -6\,g(p+1/2,q+1/2,q+1/2)+8\,g(\sqrt{3}p/2,0,0) \nn
\\& +3\,g(p,0,0) \Big]\nn,
\end{align}
where the sums extend over all positive and negative integers including zero.
(In the third and fourth terms only $p$ and $q$ are summed over,
while in the last two terms only $p$ is summed.)
Note that the time-splitting has automatically regularized the sums, and it is easy
to extract the finite part (the Epstein zeta functions $Z_3$, $Z_{3b}$, etc.\ are defined in Appendix B),
\begin{align}
E_{L}^{(D)}=\frac{1}{a}\bigg\{ & -\frac{1}{96 
\pi^2}\big[Z_3(2;1,1,1)+Z_{3b}(2;1,1,1)\big]+\frac{1}{8\pi}
\zeta(3/2)L_{-8}(3/2)+\frac{1}{16\pi} Z_{2b}(3/2;2,1)\nn
\\&-\frac{\pi}{96}-\frac{\pi\sqrt{3}}{72} \bigg\}, 
\end{align}
where (the prime means the origin is excluded) ~\cite{GlasserZucker}
 \be
 \sideset{}{'}\sum_{m,n=-\infty}^{\infty}(m^2+2 n^2)^{-s}=2\zeta(s)L_{-8}(s)\,.
 \ee
The energy then evaluates numerically to
 \be
E_{L}^{(D)} =-\frac{0.0468804266}{a}\,.
\ee	
The divergent parts, also extracted from the regularization procedure, 
follow the expected form of Weyl's law with the quartic divergence associated with the volume $V$, 
the cubic divergence associated with the surface area $S$, and the quadratic 
divergence matched with the edge coefficient
\be
E^{(D)}_{\rm div}= \frac{3 V}{2 \pi^2 
\tau^4}-\frac{S}{8\pi\tau^3}+\frac{C}			
{48\pi\tau^2}. 
\ee
Here and subsequently, the edge coefficient $C$ for a polyhedron is defined as \cite{Fedosov}
\be
C = \sum_j\left(\frac{\pi}{\alpha_j}-\frac{\alpha_j}{\pi}\right)L_j\,,
\ee
where the $\alpha_j$ are dihedral angles and the $L_j$ are the corresponding edge lengths. 
The above expression for the divergences will be the same for all subsequent cavities with 
Dirichlet boundary conditions.

\subsubsection{Neumann BC}
In the case of Neumann boundary conditions, the complete set of mode numbers must satisfy
 $0 \leq k \leq m \leq n$, excluding the case when all mode numbers are zero. 
The Neumann Casimir energy can be defined in terms of the preceding Dirichlet result as
\be
E_L^{(N)}=E_L^{(D)}-\frac{1}{8\pi a}\Big[2\,\zeta(3/2)L_{-8}(3/2)+Z_{2b}(3/2;2,1) \Big],
\ee
which gives us a numerical value of
\be
E_{L}^{(N)} =-\frac{0.1964621484}{a}\,.
\ee	
The divergent parts also match the expected Weyl terms for Neumann boundary conditions. 
We note that the cubic divergence's coefficient changes sign when comparing Dirichlet and 
Neumann divergent parts:
\be
E^{(N)}_{\rm div}=
 \frac{3 V}{2 \pi^2 \tau^4}
+\frac{S}{8\pi\tau^3}+\frac{C}{48\pi\tau^2}.
\ee
This form is also obtained for all the following calculations involving Neumann boundary conditions.
		
\subsection{Medium Tetrahedron}
\begin{figure}[h]
	\centering
	\includegraphics[scale=1, trim= 5cm 11cm 5cm 11cm, clip]{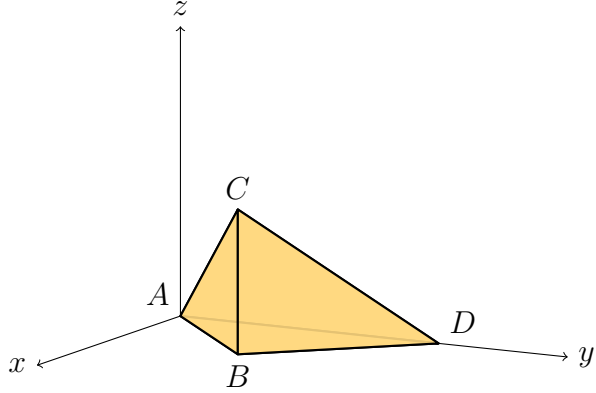}
	\caption{\label{medtetfig}Medium tetrahedron: $0<z<x$ and $x<y<2a-x$. }
\end{figure}
The eigenvalue spectrum of the medium tetrahedron,
shown in Fig.~\ref{medtetfig}, obtained by bisecting the large tetrahedron
in the $z=0$ plane, is of the same form as that of the large tetrahedron 
[Eq.~\eqref{LTeigval}] with different constraints.
\subsubsection{Dirichlet BC}
The complete set of mode numbers for the Dirichlet case satisfies the constraints $0<m<n<k<m+n$. Following the same regularization procedure used in the preceding cases, we obtain the 
Dirichlet Casimir energy in terms of the Dirichlet result for the large tetrahedron,
\begin{align}
E_{M}^{(D)} & = \frac{1}{2}\, E_{L}^{(D)}+\frac{1}{96\pi 
a}\Big[3\,\zeta(3/2)\beta(3/2)-(1+\sqrt{2})\pi^2\Big] ,
\end{align}	
where we used~\cite{GlasserZucker}	
\be
 \sideset{}{'}\sum_{m,n=-\infty}^{\infty}(m^2+n^2)^{-s}=4\zeta(s)\beta(s)\,.
\ee
The Casimir energy evaluates to
\be
E_{M}^{(D)} =-\frac{0.0799803933}{a}\,.
\ee	
Here the function $\beta$ is also defined in Appendix B.

\subsubsection{Neumann BC}
With Neumann boundary conditions, the complete set of mode numbers is restricted to 
$0 \leq m \leq n \leq k \leq m+n$, excluding the all-null case. 
As with the Dirichlet case, the Neumann Casimir energy for the medium 
tetrahedron can be expressed in terms of the Neumann result for the large tetrahedron:
\begin{align}
E_{M}^{(N)} & = \frac{1}{2}\, E_{L}^{(N)}-\frac{1}{96\pi 
a}\Big[3\,\zeta(3/2)\beta(3/2)+(1+\sqrt{2})\pi^2\Big] \nn
\\& = -\frac{0.1997008024}{a}\,.
\end{align}

\subsection{Small Tetrahedron}
\begin{figure}[h]
	\centering
	\includegraphics[scale=1, trim= 5cm 11cm 5cm 11cm, clip]{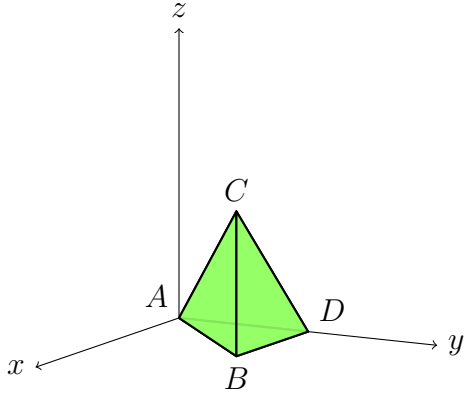}
	\caption{\label{smtetfig}Small tetrahedron:  $0<z<x$ and $x<y<a$.}
\end{figure}
The small tetrahedron (Fig.~\ref{smtetfig}) may be visualized as the result of a 
bisection of a medium tetrahedron along the plane $y=a$.  
The form of the eigenvalue spectrum for the small tetrahedron is different from the 
previous two tetrahedra but the same as the cube's\footnote{This spectrum is 
actually the same as that for the other tetrahedra,
given in Eq.~(\ref{LTeigval}), with the additional restriction that $m+n+k$ be even.}:
\be	
\lambda_{kmn}^2=\frac{\pi^2}{a^2}\left
(k^2+m^2+n^2\right)\,.
\ee
The Dirichlet case is the aforementioned ``pyramidal cavity" considered in Ref.~\cite{AhmedovSmT}.
		
\subsubsection{Dirichlet BC}
The modal restriction for the complete set is $0<k<m<n$. The finite part obtained is thus
\begin{align}
E_{S}^{(D)}=\frac{1}{a}\bigg[& 
-\frac{1}{192\pi^2}Z_3(2;1,1,1)+\frac{1}{16\pi}\zeta(3/2)L_{-8}(3/2)+\frac{1}{32\pi}\zeta(3/2)\beta(3/2)\nn
\\&-\frac{\pi}{64}-\frac{\pi\sqrt{3}}{72}-\frac{\pi\sqrt{2}}{96}\bigg],
\end{align}				
which evaluates to
\be
E_{S}^{(D)} =-\frac{0.10054146218}{a}\,.
\ee	
This result differs from that of Ref.~\cite{AhmedovSmT}. The discrepancy appears to stem from a 			
mode-counting error in Ref.~\cite{AhmedovSmT}, and the result found there is likely wrong. 
		
\subsubsection{Neumann BC}
For the Neumann case, we again find the same condition that the mode numbers 
must satisfy $0\leq k \leq m \leq n$ excluding the origin. The Neumann 
Casimir energy is derived to be
\begin{align}
E_{S}^{(N)}=E_{S}^{(D)}-\frac{1}{16\pi a}\zeta(3/2)\Big[2\,L_{-8}(3/2)
+\beta(3/2)\Big],
\end{align}		
with a numerical value of			 	
\be
E_{S}^{(N)} =-\frac{0.2587920021}{a}\,.
\ee	
		
\section{Casimir energies of Rectangular Parallelepipeds}
Amongst the geometries considered for Casimir energy calculations, 
rectangular parallelepipeds are the most straightforward. Their eigenfunctions 
and eigenvalues are well known and as such have been subject of many articles 
\cite{lukosz,lukosz2,lukosz3,ambjorn,ruggiero1,ruggiero2}. 
We rederive a few of these results in the following paragraphs. 
We consider  a generic rectangular parallelepiped of length $a$, height $b$, and width $c$.
The eigenfunctions for a Dirichlet rectangular parallelepiped are the well-known products of three 			sine functions. The spectrum is the familiar expression
\be
\lambda_{kmn}^2=\pi^2\left(\frac{k^2}{a^2}+\frac{m^2}{b^2}+\frac{n^2}{c^2}\right).
\ee

\subsection{Dirichlet BC}
The complete set of mode numbers in the Dirichlet case satisfy the restrictions 
$k>0$, $m>0$, and $n>0$ . The Casimir energy may be written in terms of Epstein 
zeta functions and the ratios: $\chi\equiv(b/a)^2$ and $\sigma\equiv(c/a)^2$
\begin{align}
\label{RectPrism}
E_{P}^{(D)} = \frac{1}{ 
a}\bigg\{&-\frac{\sqrt{\chi\,\sigma}}{32\pi^2}\,Z_3(2;1,\chi,\sigma)
+\frac{1}{64\pi}\Big[ \sqrt{\chi\,\sigma}\, Z_2(3/2;\chi,\sigma)+ \sqrt{\sigma}						
\,Z_2(3/2;1,\sigma)\nn
\\& +\sqrt{\chi} \,Z_2(3/2;1,\chi)\Big]-\frac{\pi}{96}\Big( 1
+\frac{1}{\sqrt{\chi}}+\frac{1}{\sqrt{\sigma}}\Big)\bigg\}\,.
\end{align}

\subsection{Neumann BC}
For Neumann boundary conditions the complete set is given by $k \geq 0$, $m \geq 0$, 
and $n \geq 0$, excluding the case where they are all null. 
In terms of the Dirichlet result we obtain
\begin{align}
E_{P}^{(N)} = E_{P}^{(D)}-\frac{1}{32\pi a}
\Big[\sqrt{\chi\,\sigma}\, Z_2(3/2;\chi,\sigma)+ \sqrt{\sigma}
\,Z_2(3/2;1,\sigma)+\sqrt{\chi} \,Z_2(3/2;1,\chi) \Big]\,.
\end{align}
	
\section{Casimir energies of a Cube}
The cube is a special parallelepiped with equal length, width, and height. 
Our results for the generic parallelepiped therefore apply for the particular case of 
$a=b=c$ or $\chi=\sigma=1$. This particular geometry has also been the subject of prior 
inquiries, for example Ref.~\cite{lukosz}, so we are simply rederiving these results. 
\subsection{Dirichlet BC}
The Dirichlet Casimir energy for a cube of edge length $a$ is simply
\begin{align}
E_{\rm Cube}^{(D)} 
= \frac{1}{a}\bigg[&-\frac{1}{32\pi^2}\,Z_3(2;1,1,1)
+\frac{3}{16\pi} \, \zeta(3/2)\beta(3/2)-\frac{\pi}{32}\bigg]\,,
\end{align}
from which we obtain the finite part, a result which matches that of Ref.~\cite{ambjorn}:
\be
E_{\rm Cube}^{(D)} =-\frac{0.0157321825}{a}\,.
\ee	
	
\subsection{Neumann BC}
With the same modal restrictions as for the parallelepipedal Neumann cases, the Neumann result 
can be related to the Dirichlet result with
\begin{align}
E_{\rm Cube}^{(N)} 
=E_{\rm Cube}^{(D)} -\frac{3}{8\pi a} \zeta(3/2)\beta(3/2),	
\end{align}
 which gives a numerical value already confirmed in Ref.~\cite{ambjorn}:
\be
E_{\rm Cube}^{(N)} =-\frac{0.2853094722}{a}.
\ee	

\section{Casimir energies of triangular prisms}
Since infinite triangular prisms are soluble cases~\cite{Cylpaper}, 
one would also expect finite prisms to be soluble~\cite{TerrasSwanson, Terras}. 
Indeed, one can also find the interior Casimir energies of finite triangular 
prisms of right isosceles, equilateral, and hemiequilateral cross-sections. 
The spectra differ slightly from  the infinite cases with the replacement of an
integral over longitudinal wavenumbers by a sum over discrete longitudinal eigenvalues. 

\subsection{Right Isosceles Triangular Prism}
\begin{figure}[h]
\centering
\includegraphics[scale=1, trim= 5cm 10.8cm 5cm 11cm, clip]{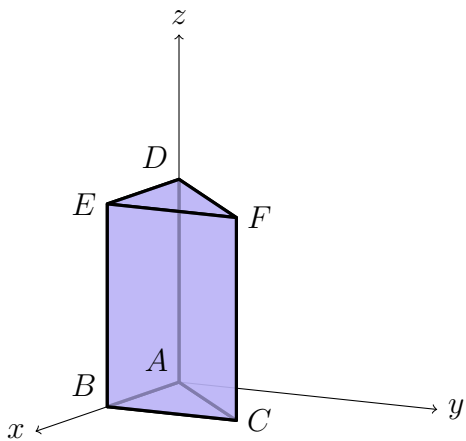}
\caption{\label{rip}Right isosceles prism.
$|DE|=|EF|=a$, $|DF|=a\sqrt{2}$, and $|BE|=b$.	}
\end{figure}
\subsubsection{Dirichlet BC}
The spectrum for a prism of right isosceles cross-section, illustrated
in Fig.~\ref{rip}, is
\be
\lambda_{kmn}^2=\frac{\pi^2}{a^2}(m^2+n^2)+\frac{\pi^2}{b^2}k^2\,.
\ee
The complete set of eigenmodes for this Dirichlet case is characterized by 
$0<n<m$ and $0<k$. The Casimir energy, in terms of $\chi \equiv (b/a)^2$, is, therefore,
\begin{align}
E_{\rm RIsoP}^{(D)}= \frac{1}{a}\bigg[&-\frac{\sqrt{\chi}}{64\pi^2}Z_3(2;1,1,\chi)
+\frac{1}{32\pi}\zeta(3/2)\beta(3/2)+\frac{\sqrt{\chi}}{64\pi}Z_2(3/2;1,\chi)
+\frac{\sqrt{\chi}}{32\pi}Z_2(3/2;1,2\chi)\nn
\\&-\frac{\pi}{64\sqrt{\chi}}-\frac{\pi(1+\sqrt{2})}{96} \bigg].
\end{align}

\subsubsection{Neumann BC}
For the Neumann case, the constraint on the mode numbers is again 
less strict, with  $0 \leq n \leq m$, $0 \leq k$ excluding $k=m=n=0$. 
In terms of the Dirichlet result, we find:
\begin{align}
E_{\rm RIsoP}^{(N)}= E_{\rm RIsoP}^{(D)}-\frac{1}{32\pi a}\bigg[2\,\zeta(3/2 
)\beta(3/2)+\sqrt{\chi}\,Z_2(3/2;1,\chi)+2\sqrt{\chi}\,Z_2(3/2;1,2\chi)\bigg].
\end{align}
		
\subsection{Equilateral Triangular Prism}
\begin{figure}[h]
\centering	
\includegraphics[scale=1, trim= 5cm 10.8cm 5cm 11cm, clip]{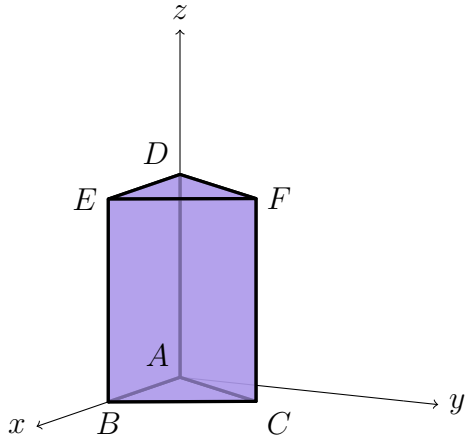}
\caption{\label{et}Equilateral prism.
$|DE|=|EF|=|FD|=a$, and $|BE|=b$. }
\end{figure}
The spectrum for an equilateral prism of height $b$, shown in Fig.~\ref{et}, is
\be
\lambda_{kmn}^2=\frac{16\pi^2}{9 a^2}(m^2-mn+n^2)+\frac{\pi^2}{b^2}k^2 \,.
\ee
		
\subsubsection{Dirichlet BC}
The constraint on $k$, $m$, and $n$ for the complete set of modes is the 
same as in the Dirichlet right isosceles case. The Casimir energy, 
in terms of $\chi \equiv (b/a)^2 $, is thus derived as
 \begin{align}
E_{\rm EqP}^{(D)}=\frac{1}{a}\Bigg\{& 
-\frac{\sqrt{3\chi}}{\pi^2}\big[Z_3(2;3,9,16\chi)+Z_{3c}(2;3,9,16\chi)\big]
+\frac{5}{48\pi}\zeta(3/2)L_{-3}(3/2)+\frac{1}{24\pi}Z_{2b}(3/2;1,3)\nn
\\&+\frac{3\sqrt{\chi}}{2\pi}Z_2(3/2;9,16\chi)-\frac{\pi}{36}-\frac{\pi}{72\sqrt{\chi}} \Bigg\},
\end{align}
where we used ~\cite{GlasserZucker}
\be
 \sideset{}{'}\sum_{m,n=-\infty}^{\infty}(m^2+3 n^2)^{-s}=2(1+2^{1-2s})\zeta(s)L_{-3}(s)\,.
\ee
This particular case was also considered earlier by Ahmedov and Duru \cite{AhmedovPrism}, although their result appears misleading.
\subsubsection{Neumann BC}
The Neumann constraint is also the same as that for the Neumann right 
isosceles triangular prism. 
Similarly to previous cases, we relate the Neumann result to the Dirichlet result,
\begin{align}
E_{\rm EqP}^{(N)}=E_{\rm EqP}^{(D)}-\frac{1}{24\pi 
a}\Big[5\,\zeta(3/2)L_{-3}(3/2)+2\,Z_{2b}(3/2;1,3)+72\sqrt{\chi}\,\,Z_2(3/2;9,16\chi)\Big].
\end{align}

\subsection{Hemiequilateral Triangular Prism}
\begin{figure}[h]
\centering
\includegraphics[scale=1, trim= 5cm 10.8cm 5cm 11cm, clip]{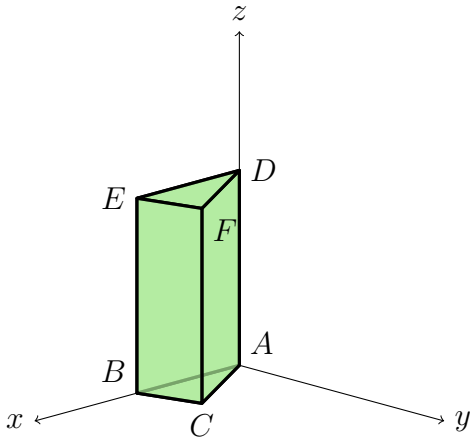}
\caption{\label{hep}Hemiequilateral prism.
$|DF|=a\sqrt{3/4}$, $|EF|=a/2$, $|DE|=a$, and $|BE|=b$.}
	\end{figure}
The hemiequilateral triangular prism (Fig.~\ref{hep}) or prism with
cross-section being a triangle with angles
$(\pi/2,\,\pi/3,\,\pi/6)$ shares the same spectral form as the equilateral triangular prism,
\be
\lambda_{kmn}^2=\frac{16\pi^2}{9 a^2}(m^2-mn+n^2)+\frac{\pi^2}{b^2}k^2\,.
\ee
They differ, however, in the constraints for each boundary condition.
\subsubsection{Dirichlet BC}
The complete set of modes must satisfy $0<k$, and $0<n<m<2n$. Again, in terms of  
$\chi \equiv (b/a)^2 $, we find that the Dirichlet Casimir energy is of the form
\begin{align}
E_{\rm HemP}^{(D)}=\frac{1}{2} E_{\rm EqP}^{(D)}
+\frac{\sqrt{3\chi}}{4\pi a}Z_2(3/2;3,16\chi)
-\frac{\pi}{72 a}\Big(\sqrt{3}+\frac{3}{4\sqrt{\chi}}\Big).
\end{align}
		
\subsubsection{Neumann BC}
The completeness constraint for the Neumann case is again less strict than for 
the Dirichlet case with $0 \leq k$, $0 \leq n \leq m \leq 2n$ excluding the origin. 
In relation to the previous result, we write
\begin{align}
E_{\rm HemP}^{(N)}=\frac{1}{2} E_{\rm EqP}^{(N)}
-\frac{\sqrt{3\chi}}{4\pi a}Z_2(3/2;3,16\chi)-\frac{\pi}{72 
a}\Big(\sqrt{3}+\frac{3}{4\sqrt{\chi}}\Big)\,.
\end{align}
	 
\section{Casimir energies of a Sphere}
The sphere is also one of the geometries most often the subject of 
Casimir energy calculations
(For more complete references see Ref.~\cite{milton-rev}.) 
We report results found in the literature for a 
sphere of radius $a$ satisfying Dirichlet and Neumann boundary conditions. 
A noteworthy difference between the calculations of the energies for tetrahedra and prisms as compared
to a sphere is that in the polyhedral cases only the interior modes are considered 
(the exterior modes are unknown) whereas in the spherical case both interior and exterior are 
(necessarily) included to cancel the curvature divergences.

\subsection{Dirichlet BC}
The Dirichlet Casimir energy of a sphere is well known and may be found in 
Ref.~\cite{DirichletSphere},
\be
E_{\rm Sphere}^{(D)} =\frac{0.0028168}{a}.
\ee	
\subsection{Neumann BC}
The Neumann result is also well known and can be found in Ref.~\cite{Nesterenko},
\be
E_{\rm Sphere}^{(N)} =-\frac{0.223777}{a}.
\ee	

\section{Systematics of Casimir energies}
\begin{figure}[!h]
\begin{center}
\includegraphics[scale=1
]{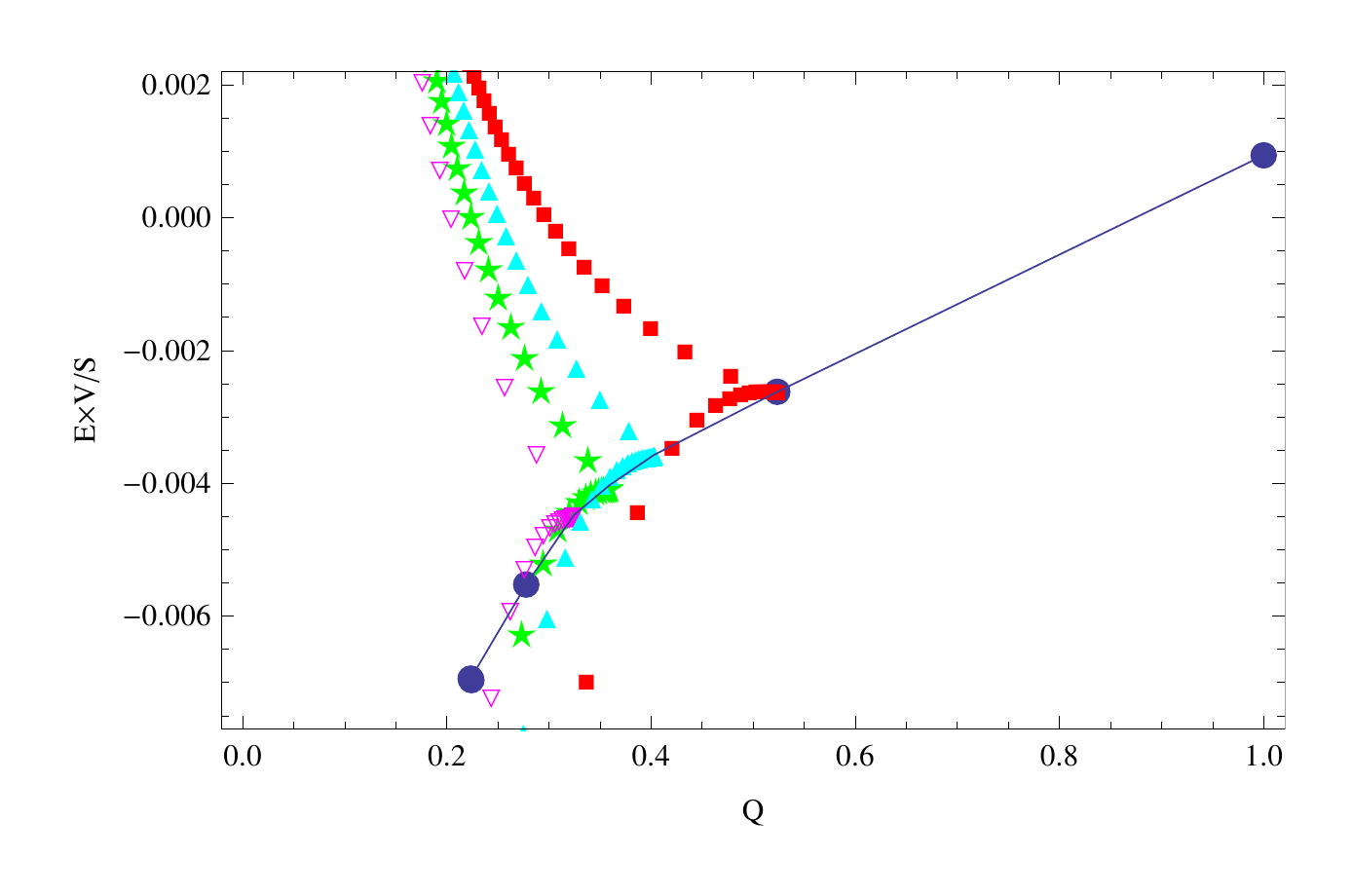}
\end{center}
\caption{Scaled Dirichlet energies vs.\ isoareal quotients. From left 
to right,  the circular markers indicate results for the small and 
medium tetrahedra (which cannot be resolved on this graph), 
the large tetrahedron, the cube, and the sphere. The square,  filled triangle, 
star, and empty triangle markers correspond respectively to square prisms 
(parallelepipeds with $\sigma=1$), and equilateral,
right isosceles,  and hemiequilateral triangular prisms, respectively.
The prism energies become more negative for $b\to 0$, less negative for 
$b\to\infty$.  Note that the cusps, corresponding to the maximal isoareal 
coefficient for a given class of cylinders, also lie close to a universal 
curve that passes through the tetrahedral points.}
\label{plot2dD}
\end{figure}

\begin{figure}[!h]
\begin{center}
\includegraphics[scale=0.7
]{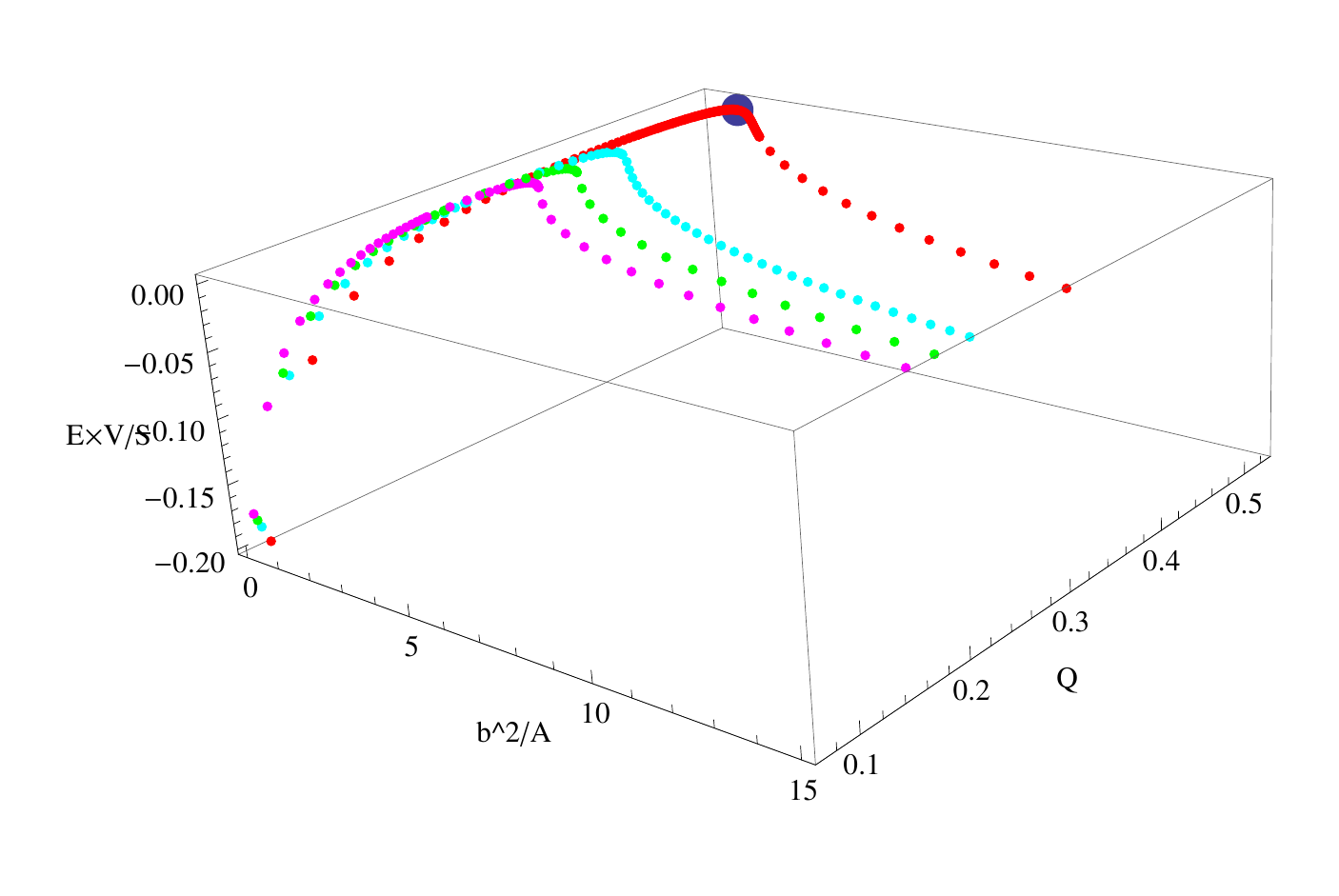}
\end{center}
\caption{Scaled Dirichlet energies of prisms vs.\  
isoareal quotients $\mathcal{Q}$ and $b^2/A$, where $A$ is the cross-sectional
area.  Starting from the lowest $\mathcal{Q}$-values, the curves correspond 
respectively to hemiequilateral,  right isosceles, and equilateral 
triangular prisms, and square prisms (parallelepipeds with $\sigma=1$). 
The square prisms' curve goes through the cube's data point 
(displayed prominently).}
\label{plotD3d}
\end{figure}

\begin{figure}[!h]
\begin{center}
\includegraphics[scale=1
]{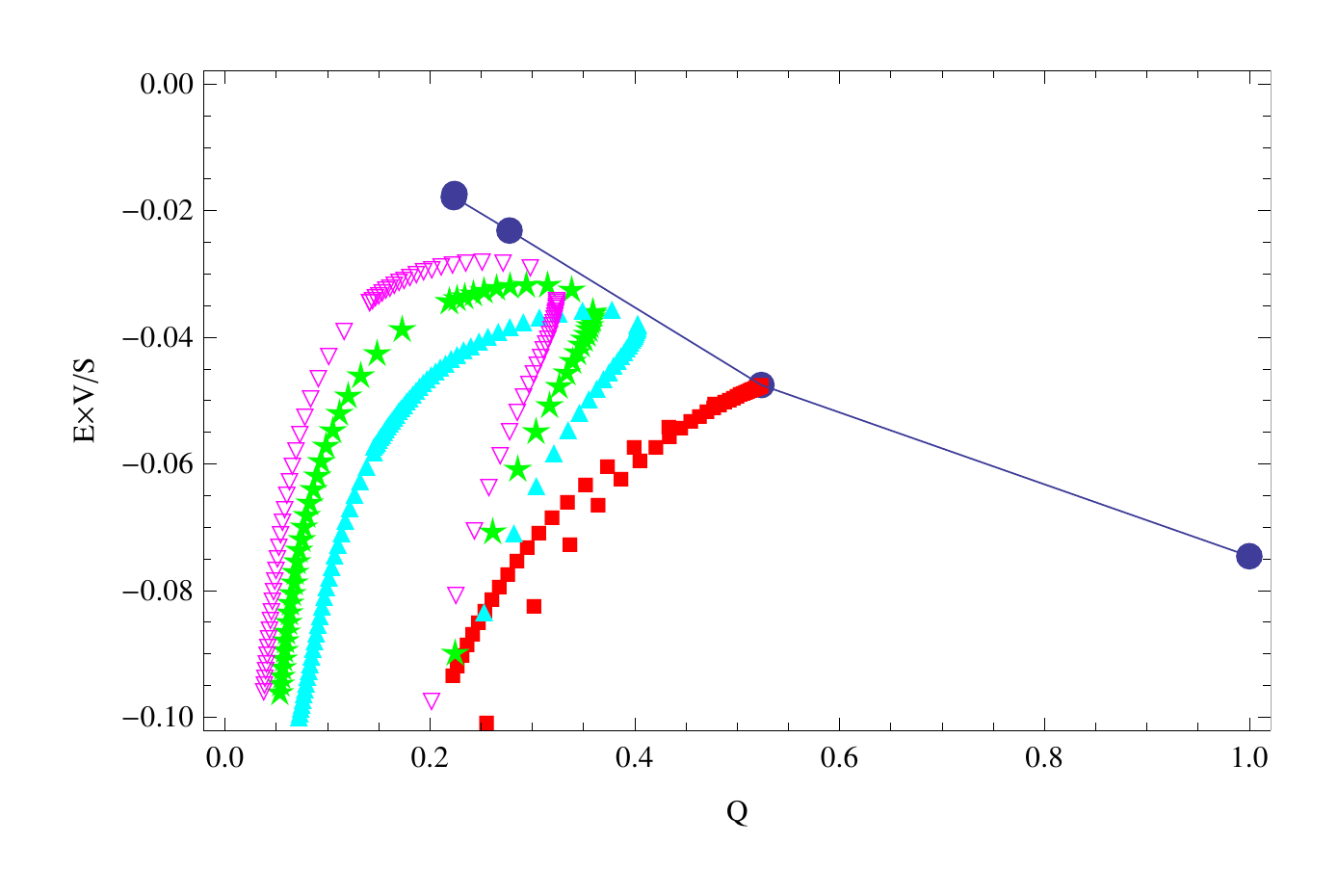}
\end{center}
\caption{Scaled Neumann energies vs.\ isoareal quotients. From left
 to right,  the circular markers indicate results for the small and
medium tetrahedra (which cannot be resolved on this graph),
the large tetrahedron, the cube, and the sphere. The square,  filled triangle,
star, and empty triangle markers correspond respectively to square prisms
(parallelepipeds with $\sigma=1$), and equilateral, right isosceles,  and 
hemiequilateral triangular prisms, respectively. The energies in this
case are always negative. Note that the cusps, corresponding to the 
maximal isoareal coefficient for a given class of cylinders, again 
lie close to a universal curve that passes through the tetrahedral points.}
\label{plot2dN}
\end{figure}

As indicated in the introduction, the relation between the self-energy of 
a system 
and its geometry is not obvious. (We set aside the more 
serious physical difficulty
as to the meaning and the observability of Casimir self-energies.)
Having additional data, such as the self-energies of tetrahedra 
and finite prisms, 
may help in shedding some light on this problem. 
Our analysis is similar to the one applied 
earlier to infinite prisms~\cite{Cylpaper}.
In terms of the volume $V$ and the surface area $S$ of the bodies,
the dimensionless scaled Casimir energies, 
$E_{\rm Sc}=E\times V/S$, are tabulated 
in Table \ref{tab1}, and
are plotted against the corresponding isoareal quotients, 
$\mathcal{Q}=36 \pi V^2/S^3$ in Fig.~\ref{plot2dD}. 
 It is also possible to look at the prisms in a more revealing light by 
plotting their scaled energies with respect to the parameters 
$\mathcal{Q}$ and $b^2/A$ (Fig. \ref{plotD3d}).
The corresponding results for Neumann energies are given in Fig.~\ref{plot2dN}.
 \begin{table}[h]
 \centering
 \begin{tabular}{l @{\quad\quad}c @{\quad}c @{\quad}c}
\hline\noalign{\smallskip}
Cavity &  $\mathcal{Q}$ &  $E_{\rm Sc}^{(D)}$ &  $E_{\rm Sc}^{(N)}$ \\
\hline\noalign{\smallskip}
Small T.      & 0.22327     &  $-0.00694$ & $-0.01787$ \\
Medium T.  & 0.22395 & $-0.00696$ & $-0.01739$\\
Large T.   & 0.27768  & $-0.00552$ & $-0.02315$\\
Cube   & 0.52359  & $-0.00262$ & $-0.04755$\\
Spherical Shell & 1 & \hphantom{$-$}0.00093 & $-0.07459$\\
\hline
\end{tabular}
\caption{\label{tab1}Scaled energies and isoareal quotients.}
\end{table}

\section{Limiting cases of prisms}
For the prisms we considered, with square and equilateral, hemiequilateral, 
and right-isosceles triangular cross sections, we can examine two limits 
which reduce to known cases.  With $b$ as the height of the prism, the 
limit $b\to0$ must coincide with the classic Casimir case of parallel plates
of area $A$, 
\be
E\to-\frac{\pi^2}{1440 b^3} A.
\ee
And with $b\to\infty$ we must recover the energy for infinite cylinders given 
in Ref.~\cite{Cylpaper},
\be
E\to \mathcal{E}_{\rm cyl} b ,
\ee
in terms of the energy/length for the cylinder, $\mathcal{E}_{\rm cyl}$.
For a prism of cross section $A$ and cross-sectional perimeter $P$, 
the scaled Casimir energy is
\be
E_{\rm Sc}=E\frac{V}S=\frac{EAb}{(2A+Pb)},
\ee
where $S$ denotes the total surface area of the prism and $V$ its volume.
Thus, in the limit of vanishing height,
\be
b\to 0:\quad E_{\rm Sc}\to -\frac{\pi^2}{2880}\frac{A}{b^2}.
\ee
In the limit of infinite length,
\be
b\to\infty:\quad E_{\rm Sc}\to (\mathcal{E}_{\rm cyl} b)\frac{A}{P}.
\label{infPrism}
\ee
These limits are exact.

\begin{figure}[!h]
\centering
\subfloat[Subfigure 1 list of figures text][$b/a \to 0$ limit.]{
\includegraphics[scale=0.8,width=0.45\textwidth]{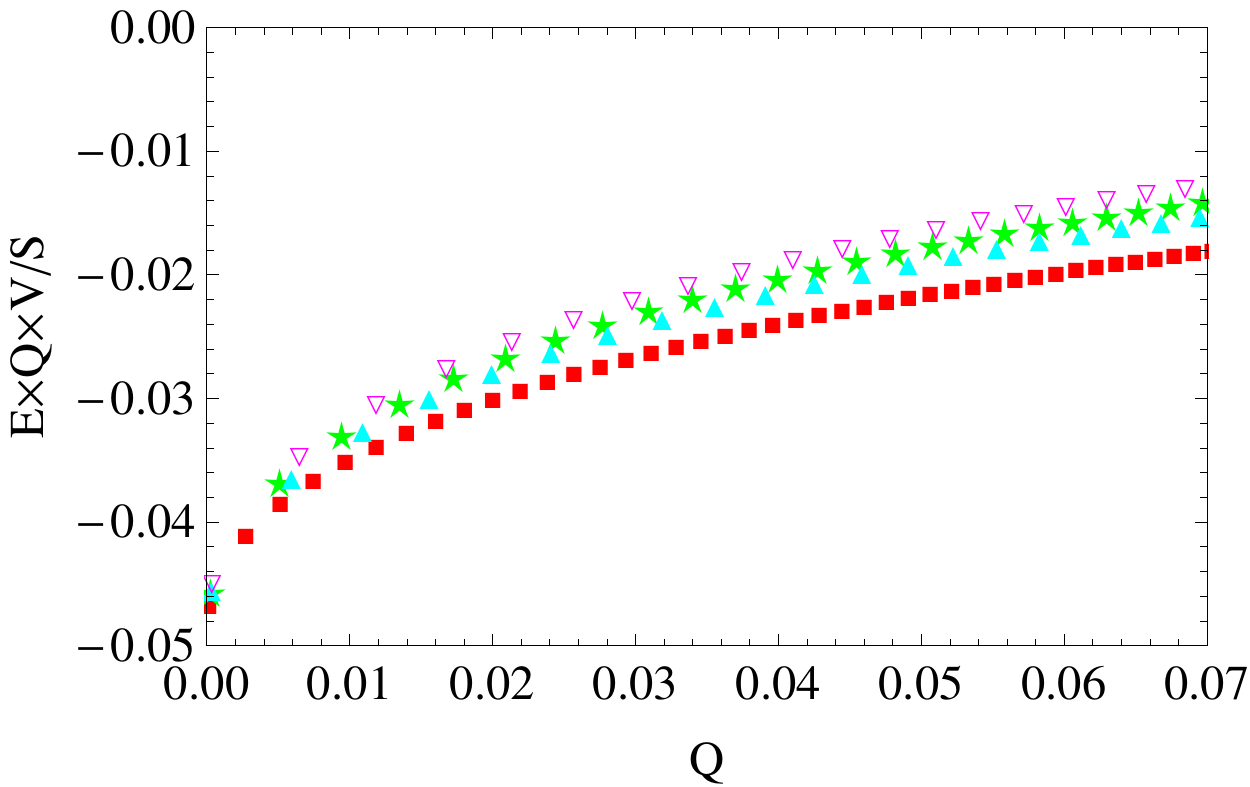}
\label{plSmall}}
\subfloat[Subfigure 2 list of figures text][$b/a\to\infty$ limit.]{
\includegraphics[scale=0.8,width=0.45\textwidth, trim= 0cm 0.3cm 0cm 0cm, clip ]{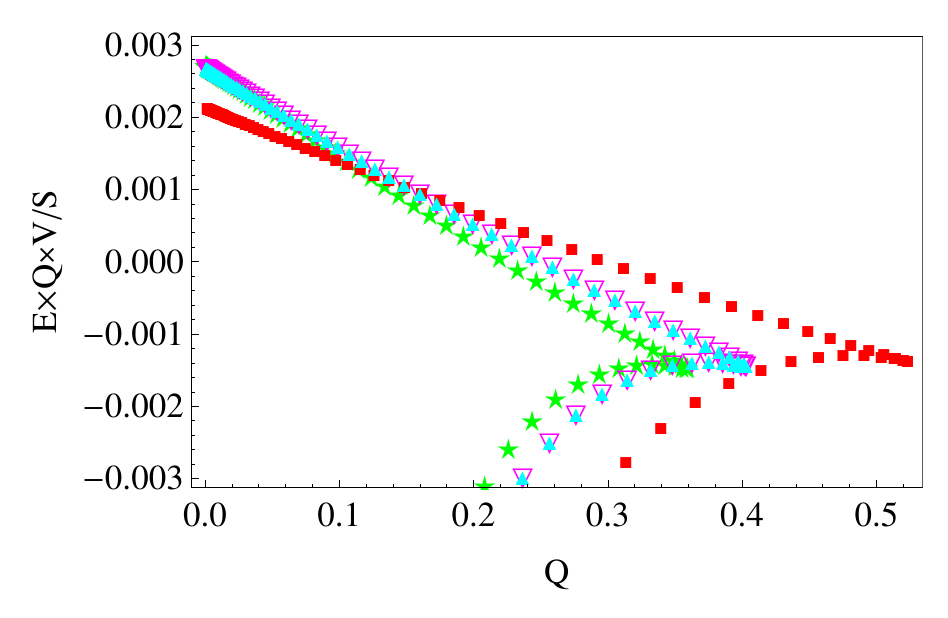}
\label{plLarge}}
\caption{$E\mathcal{Q}V/S$ plotted versus $\mathcal{Q}$ in the
$b/a \to 0$ (left panel) and $b/a\to\infty$ (right panel)
limits for  Dirichlet prisms. The curves 
consisting of square, filled triangle, star,  and empty triangle markers 
correspond respectively to prisms of square, and equilateral,  
right isosceles, and hemiequilateral triangular cross-sections. 
The curves  converge to the expected value of 
$-\pi^3/640=-0.0484473$ as $b/a\to 0$. In the $b/a\to \infty$ limit, they converge respectively to the values of $0.00213$, $0.00269$, $0.00274$, and $0.00277$ which are simply obtained from Eq.~(\ref{infPrism}). These limits are not particularly close to the proximity force approximation value of $\pi^3/10240=0.00303$.}
\label{figxx}
\end{figure}

It may be interesting to express these limits in terms of the isoareal
quotients:  
\be
\mathcal{Q}=\frac{36\pi V^2}{S^3}=\frac{36 \pi b^2 A^2}{(2A+bP)^3},
\ee
whence,
\be
b\to 0:\quad E_{\rm Sc}\to -\frac{\pi^3}{640 \mathcal{Q}}.
\ee
In this limit $\mathcal{Q}\to0$.  A similar expression may be obtained
in the large $b$ limit if we use the proximity force approximation
as discussed in Ref.~\cite{Cylpaper}:
\be
\mathcal{E}_{\rm cyl}A\approx \frac{\pi^2}{368640}\left(\frac{P^2}A\right)^2,
\ee
which becomes exact as the smallest angle of the triangle vanishes.
In this approximation
\be
b\to\infty:\quad E_{\rm Sc}\approx\frac{\pi^3}{10240}\frac1{\mathcal{Q}},
\ee
where again $\mathcal{Q}\to 0$.
In Fig.~\ref{figxx} we show how the long and short prism limit are
approached by our data.

\begin{figure}[h]
\centering
\subfloat[Subfigure 1 list of figures text][$b/a \to 0$ limit.]{
\includegraphics[scale=0.8,width=0.45\textwidth, trim= 0.55cm 0.55cm 0.45cm 0.55cm, clip]{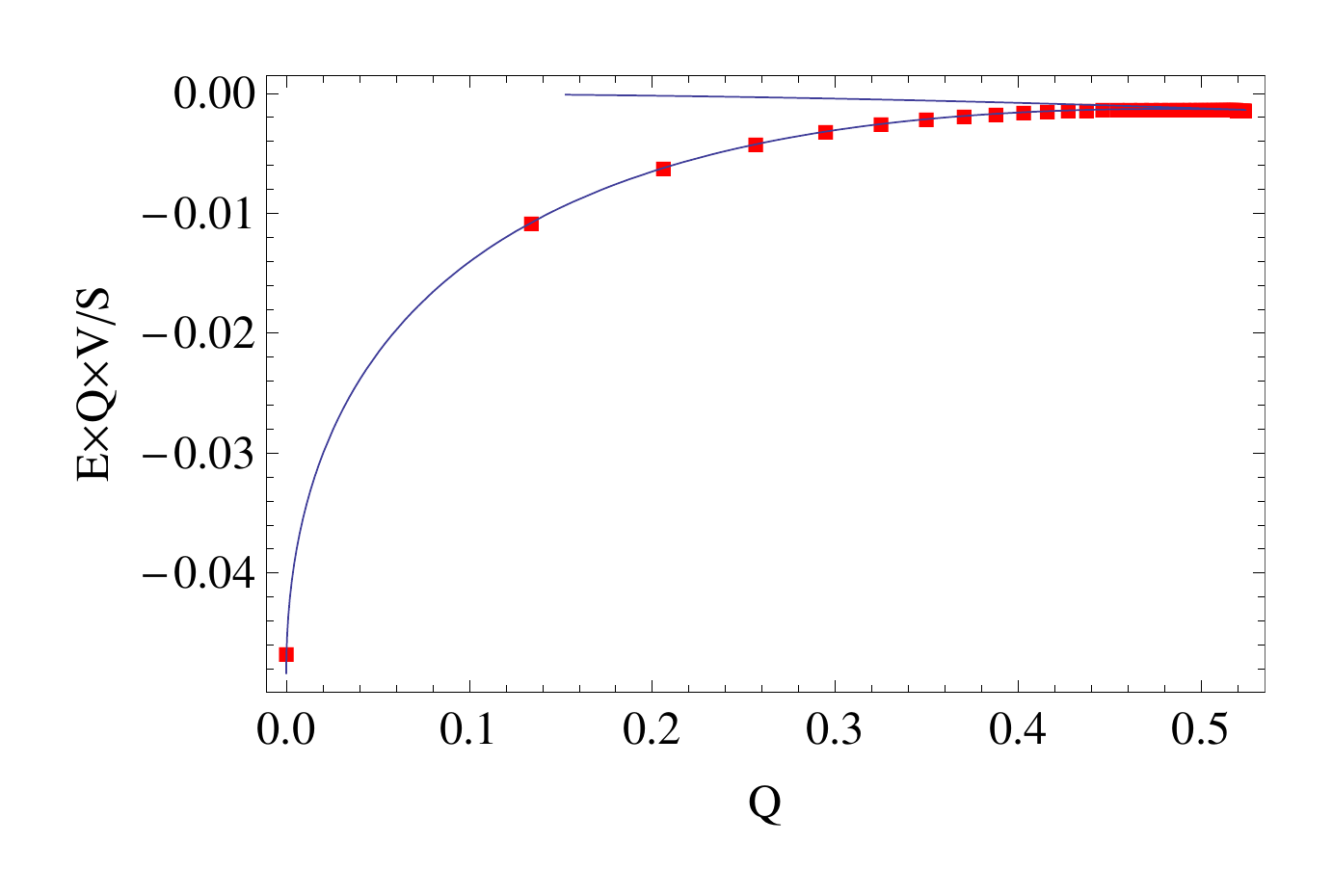}
\label{fig:SqSmallb}}
\small
\subfloat[Subfigure 2 list of figures text][$b/a\to\infty$ limit.]{
\includegraphics[scale=0.8,width=0.45\textwidth, trim= 0.55cm 0.55cm 0.45cm 0.55cm, clip]{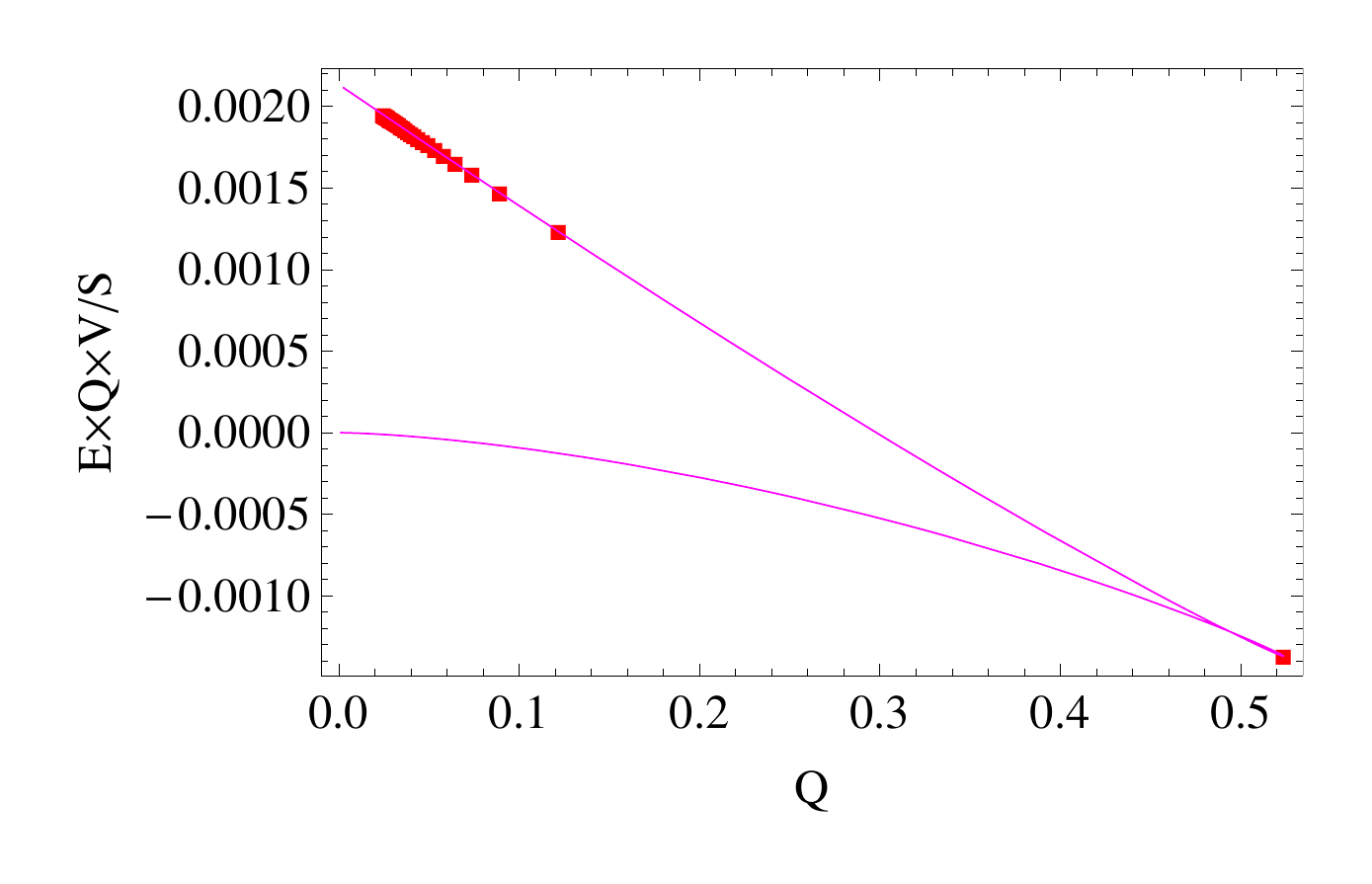}
\label{fig:SqLargeb}}
\caption{\label{btozero}Asymptotic fits for square prisms.  The small
height ($b/a\to0$) (left panel) and large height ($b/a\to\infty$) (right panel)
asymptotic energies are shown,
corresponding to Eqs.~(\ref{short}) and (\ref{long}), respectively.
Plotted is the energy multiplied by $\mathcal Q V/S$, as a function of
$\mathcal{Q}$.  The square markers show the exact results, which lie
very close to the physical branch in each case.}
\end{figure}

Let us examine the $b\to0$ limit for the example of the square prism, where the
Dirichlet energy is given by Eq.~(\ref{RectPrism}), with $\sigma=1$.  
Using the Euler-Maclaurin summation formula, we find the asymptotic limit 
of that expression for short prisms to be
\be \label{short}
E_{\rm Sq}^D\sim -\frac{\pi^2a^2}{1440 b^3}\left[1-\frac{90}{\pi^3}
\frac{b}a\zeta(3)+\frac{15}{\pi}\frac{b^2}{a^2}\right], \quad b/a\to 0.
\ee
There are only exponentially small corrections to this result.  
Similarly in the long distance (infinitely long cylinder) limit, $b\gg a=c$,
\be\label{long}
E_{\rm Sq}^D\sim -\frac1{16\pi a}\left\{\frac{b}a\left[\frac{\pi}3 G
-\zeta(3)\right]-
\left[\zeta(3/2)\beta(3/2)-\frac{\pi^2}3\right]\right\},\quad a/b\to0,
\ee
again, up to exponentially small terms. (Here $G$ is Catalan's constant.) 
Note that the leading terms in the expressions are the correct limiting forms: 
that in Eq.~(\ref{short}) is Casimir's result for plates of area $a^2$, and 
that in Eq.~(\ref{long}) is that for a square cylinder found in Eq.~(5.4) of 
Ref.~\cite{Cylpaper}.  In Fig.~\ref{btozero} we plot, parametrically, these 
asymptotic limits
against the exact form of the isoareal
coefficient, e.g. $\mathcal{Q}=9\pi b^2a/2(a+2b)^3$ for the square prism.
We note that both limiting forms are well reproduced by the 
data for the prisms, all the way down to the cusp which occurs for the maximal 
value of the isoareal coefficient.  In fact, the accuracy of the asymptotic
formulas is remarkable: At the cusp, $b=a$, Eq.~(\ref{short}) gives for the
scaled energy of the cube the value $-0.00261078$, while Eq.~(\ref{long})
gives $-0.00261479$, both differing by less than $0.5\%$ from the exact value
$-0.00262203$.

\section{Conclusions}
In this paper, we have extended the work of Ref.~\cite{Cylpaper} from 
infinite cylinders to finite prisms and to integrable tetrahedra.  
The previous work was essentially two dimensional,
so it was possible to give closed form results for the Casimir self energy. 
This is  apparently not possible, at least not currently, for the cases 
considered in this paper.
Nevertheless, our answers are expressed in terms of Epstein zeta functions
and other well-known functions, so numerical results of arbitrary accuracy are 
available.

The emerging systematics are very intriguing: Not only do the three integrable 
tetrahedra and the cube and the sphere seem to lie very close to a universal 
curve (there is a very slight discrepancy in the case of the small/medium 
tetrahedra) but the maximal isoareal limits of the triangular prisms line up 
as well. The square prisms are well described by the asymptotic formul\ae\
(\ref{short}) and (\ref{long}), and similar formul\ae\ exist for the other
prisms. These results are  not yet conclusive, since the cases we can evaluate
are limited.  Numerical work will have to be done to explore the geometrical 
dependence of the Casimir energy of cavities composed of flat surfaces of 
arbitrary shape.

The reader may rightly inquire as to what the physical significance
of these results may be.  Self-energies are inherently resistant to
physical observation: they describe the energy required to assemble the
configuration, but not the energy required to remove one side of a box,
for example.  And, here, the difficulty of interpretation is somewhat compounded, since we are unable to include effects of the modes exterior
to the cavity.  For the case of a sphere or a cylinder, it is inconsistent
not to do so, since a unique finite result cannot be obtained except for
a shell of infinitesimal thickness, with both interior and exterior
contributions.  Here, however, we are considering cavities with flat sides,
so an unambiguous finite part may be extracted, with the volume, surface,
and corner Weyl divergences uniquely removable.  Curvature divergences
correspond to a logarithmic divergence in the energy, so they introduce
an arbitrary scale, and there is no meaning to interior and exterior mode
contributions separately.

 The fact that these interior energies are rigorously computable and  exude a sense of overall order is already an important step. One could very well rephrase this problem in terms of cavities in conducting materials, where the exterior would not be of significance. In fact, we argue that one would only need to compute the interior energies of arbitrary domains with planar boundaries to observe significant patterns. Even then, the converse problem for domains with smooth boundaries in conducting materials would still arise. It is very likely the solution lies in the transition from planar to smooth boundaries, from discreteness to continuity. Nevertheless, it is remarkable, even though it appears fortuitous,
that our results for tetrahedra and prisms appear to lie on a curve which
intercepts the interior plus exterior Casimir energy for a sphere. The answers are most surely rooted in what happens in between, and the appearance of the curvature-related logarithmic terms. The zero-point energies for spheres are considered purely in the spirit of future analyses.

At the very least, the work reported here is of mathematical interest in elucidating
the systematics of Casimir energies.  
In Ref.~\cite{DirichletSphere}, we explored
the systematic dependence on dimension for hyperspheres.  
Here, we have discovered some remarkable systematic behavior,
where the values of Casimir energies vary smoothly with geometrical
parameters. Understanding such systematics is vital for future developments
involving quantum vacuum effects, which will undoubtedly yield applications
in nanoscience \cite{binns}.

In addition to the worthwhile issues raised in the previous paragraphs, work on other boundary conditions, in particular electromagnetic boundary 
conditions, is currently under way.
Unlike for cylinders, the electromagnetic 
energy of a tetrahedron is not merely the sum of Dirichlet and Neumann parts; 
there is no break-up into TE and TM modes in general.  So this is a formidable 
task.  Higher-dimensional analogues, polytopes, are also currently the subject 
of ongoing work.
\acknowledgments
We thank the US National Science Foundation and the US Department of Energy for
 partial support of this work.  We further thank Nima Pourtolami and Prachi
Parashar for collaborative assistance.

\appendix
\section{Poisson Resummation Formulae}
\label{appendixA}

We consider the Poisson resummation of the traced cylinder kernel 
of an arbitrary real quadratic form,
\be 
\mathcal{S}=\sum_{m_1,\ldots,m_n=-\infty}^{\infty}e^{-\tau \sqrt{(m+a)_j\, A_{jk}\, 
(m+a)_k}}\,.
\ee
Taking the Fourier transform of the summand of $\mathcal{S}$ 
and using Eq.~(\ref{psf}) gives
\be
\mathcal{S}=\sum_{m_1,\ldots,m_n=-\infty}^{\infty}\int_{-\infty}^{\infty}\prod_{j=1}^n \, du_j \, 
e^{2\pi i u_j m_j} e^{-\tau \sqrt{(u_i+a_i)A_{ik}(u_k+a_k)}}\,.
\ee
We shift the variables
\be
u_j \rightarrow u_j-a_j \,,\ee
and diagonalize $A$
\be
B_{ij}=U_{ik}\,A_{km}\,U_{mj}^{\mathsf{T}}\,.\ee
A redefinition of the integration variables follows,
\be
v_j=U_{jk} u_k\,,
\ee
as well as the summation variables,
\be
q_j=U_{jk}\,m_k\,.
\ee
As a result of these transformations, we recognize that the Jacobian of the 
transformation matrix is unity,
\be
\prod_{j=1}^n du_j=\prod_{j=1}^n dv_j\,.\ee
We are now ready to change to hyperspherical coordinates. First, we define 
\be
R_j=\sqrt{B_{jj}}\,v_j
 \ee
and
\be
k_j=\frac{q_j}{\sqrt{B_{jj}}}
\ee
which allows us to write
\be
v_j\,q_j=kR\cos{\theta}\,.
\ee
Effectuating the change of variables gives us
\be\prod_{j=1}^n dv_j=\left|\det\left(B\right)\right|^{-1/2}R^{n-1}dR\,d\phi\,
\left(\sin{\theta}\right)^{n-2}
d\theta\,\prod_{j=1}^{n-3}\left(\sin{\theta_j}\right)^{j}\,d\theta_j\,.\ee
The $\phi$-integral produces a $2\pi$ and the integrals for the first $(n-3)$ 
$\theta_j$ angles give
\be\prod_{j=1}^{n-3}\left(\int_{0}^\pi \sin^j\theta\,d\theta\right)=
\frac{\pi^{(n-3)/2}}{\Gamma{\left((n-1)/2\right)}}\,.\ee
We are now able to focus on the remaining $\theta$-integral,
\be
\int_{0}^\pi\left(\sin\theta\right)^{n-2}e^{2\pi i k R\cos\theta}\,d\theta=
\pi^{(3-n)/2}\Gamma{\left((n-1)/2\right)}\left(k R\right)^{(2-n)/2}
J_{(n-2)/2}\left(2\pi k R\right)\,.\ee
The last integral, the $R$-integral, is evaluated rather straightforwardly,
\be
\int_{0}^\infty dR\,R^{n/2} J_{(n-2)/2}\left(2\pi k R\right)e^{-\tau R}
=\frac{\tau\, 2^{n-1}\pi^{(n-3)/2}k^{(n-2)/2}\Gamma\left((n-1)/2\right)}{\left(\tau^2+4\pi^2 k_j\,k_j\right)^{(n+1)/2}}\,,
\ee
and putting everything together we obtain:
\be 
\mathcal{S}=\frac{2^n \pi^{(n-1)/2}\,\Gamma((n+1)/2)}{\left|
\det\left(A\right)\right|^{1/2}}\sum_{m_1,\ldots,m_n=-\infty}^{\infty}
\frac{\tau\,e^{-2 \pi i\, m_j\,a_j}}{
\left(\tau^2+4\pi^2k_j\,k_j\right)^{(n+1)/2}}.
\ee
From this result we obtain the
following resummed expressions we use in the paper:

\begin{align}
\bigg(-\frac{d}{d\tau}\bigg)\sum_{p,q,r=-\infty}^{\infty}
e^{-\tau\sqrt{\alpha(p+a)^2+\beta(q+b)^2+\gamma(r+c)^2}}
= &\frac{24\pi}{\sqrt{\alpha\beta\gamma}\,\tau^4}
-\frac{1}{2\pi^3\sqrt{\alpha\beta\gamma}}\,
\\&\times\sideset{}{'}
\sum_{p,q,r=-\infty}^{\infty}\bigg(
\frac{e^{-2\pi i(pa+qb+rc)}}{(p^2/\alpha + q^2/\beta + r^2/\gamma)^2}\bigg),\nn
\end{align}

\be
\bigg(-\frac{d}{d\tau}\bigg)\sum_{p,q=-\infty}^{\infty}
e^{-\tau\sqrt{\alpha(p+a)^2+\beta(q+b)^2}}
=\frac{4\pi}{\sqrt{\alpha\beta}\,\tau^3}
-\frac{1}{4\pi^2\sqrt{\alpha\beta}}\,\sideset{}{'}\sum_{p,q=-\infty}^{\infty}
\frac{e^{-2\pi i(pa+qb)}}{(p^2/\alpha + q^2/\beta)^{3/2}},
\ee
\be
\bigg(-\frac{d}{d\tau}\bigg)\sum_{p=-\infty}^{\infty}
e^{-\tau\sqrt{\alpha(p+a)^2}}=\frac{2}{\sqrt{\alpha}\,\tau^2}
-\frac{\sqrt{\alpha}}{2\pi^2}\,\sideset{}{'}\sum_{p=-\infty}^{\infty}\frac{e^{-2\pi i (pa)}}
{p^2}.
\ee
Here the prime means that all positive and negative integers are included
in the sum, but not the case where all the integers are zero.

\section{Epstein Zeta Functions}\label{appendixB}
We define the following Epstein zeta functions: 
\be
Z_3(s;a,b,c)=\sideset{}{'}\sum_{k,m,n=-\infty}^{\infty}(a\,k^2+b\,m^2+c\,n^2)^{-s},
\ee
\be
Z_{3b}(s;a,b,c)=\sideset{}{'}
\sum_{k,m,n=-\infty}^{\infty}(-1)^{k+m+n}(a\,k^2+b\,m^2+c\,n^2)^{-s},
\ee
\be
Z_{3c}(s;a,b,c)=\sideset{}{'}\sum_{k,m,n=-\infty}^{\infty}(-1)^{k+m}(a\,k^2+b\,m^2+c\,n^2)^{-s},
\ee
\be
	Z_{2b}(s;a,b)=\sideset{}{'}\sum_{m,n=-\infty}^{\infty}(-1)^{m+n}(a\,m^2+b\,n^2)^{-s}.
\ee
Here, sums are over all integers, positive, negative, and zero, excluding the single
point where all are zero.
They are summed numerically using Ewald's method~\cite{GlasserZucker, Crandall1, Crandall2}. A few specific values needed for calculations:
\be
Z_3(2;1,1,1)=16.5323159598,
\ee
\be
Z_{3b}(2;1,1,1)=-3.8631638072,
\ee
\be
Z_{3c}(2;1,1,1)=-1.8973804658,
\ee
\be
Z_{2b}(3/2;1,2)=-1.9367356117,
\ee
\be
Z_{2b}(3/2;1,3)=-1.8390292892.
\ee
The Dirichlet $L$-series are defined as $L_k(s)= \sum_{n=1}^{\infty}\chi_k(n)\,n^{-s}$ where $\chi_{k}$ 
is the number-theoretic character \cite{GlasserZucker}. The Dirichlet beta function, also known as $L_{-4}$, is usually defined as $\beta(s)=\sum_{n=0}^{\infty}(-1)^{n}(2n+1)^{-s}$.

\end{document}